\documentclass[12pt,twocolumn]{article}
\usepackage{widetext}
\usepackage{flushend}
\usepackage{cuted}
\usepackage{color}
\usepackage{graphicx}
\usepackage{hyperref}
\usepackage{fancyhdr}
\usepackage[raggedright]{titlesec}
\begin{document}
\title{Doppler Shift of the de Broglie Waves- Some New Results from
Very Old Concepts\\
\vspace{.3cm}
\large{ 
Sanchari De and Somenath Chakrabarty}\\
\vspace{0.3cm}
\normalsize{
Department of Physics, Visva-Bharati, Santiniketan 731 235, 
India\\
{\small somenath.chakrabarty@visva-bharati.ac.in}\\
\vspace{0.3cm}
(Submitted dd-mm-yyyy)}\\
\rule[0.1cm]{16cm}{0.02cm} \\
\textbf{Abstract}\\
\flushleft \normalsize {
The Doppler shift of de Broglie wave is obtained for fermions and
massive bosons 
using the conventional form of Lorentz transformations for momentum and
energy of the particles. A formalism is developed to obtain the variation of
wave length for de Broglie waves with temperature for individual
particles using the classic idea of Wien in a many body Fermi gas or
massive Bose gas. 
} \\
\rule[0.1cm]{16cm}{0.02cm} 
}
\date{}
\maketitle
\thispagestyle{fancy} 
\lhead{\textbf{Physics Education}}
\chead{\thepage}
\rhead{\bf {dateline}(to be added by Editor)}
\lfoot{Volume xx, Number y Article Number : n.(to be added by editor) }
\cfoot{ }
\rfoot{www.physedu.in}
\renewcommand{\headrulewidth}{0.4pt}
\renewcommand{\footrulewidth}{0.4pt}
\section{Introduction}
It is well known that the pitch
of an audible acoustic wave appears to change if there is a relative
motion between the source and the observer. Such apparent change in
frequency of sound waves when there is a non-zero relative velocity is known as the Doppler
effect in the case of acoustics. Same kind of physical effect
is also been observed in the case of optics with an apparent
change in wavelength or color of the light. With simple theoretical
calculation one can show that the effect is common to all kinds of travelling
waves. However, to make appreciable change in frequency or wavelength for light waves
the relative velocity must be a considerable fraction of the velocity of light. Since the speed of light is
extremely high compared to 
the speed of any terrestrial object, it is difficult to detect any
measurable or observable change in wavelength or color of the emitted
light from any source having relative motion with respect to the
observer. 
Whereas many of the stellar objects are moving with very high
velocity, therefore wavelength of
light emitted from these heavenly bodies, moving either towards
or away from the earth show measurable amount of blue or redshift respectively in
the spectral lines. The possibility of shift in the position of a
spectral line due to relative motion of the source and the observer
was first pointed out by Doppler in the year 1842. From the
observed red shift and blue shift of spectral lines, it has been
reported that the stars like Sirius, Caster, Regulus are moving away
from the earth, whereas, the stars like Arcturus, Vega and $\alpha$
Cygni are moving towards the earth. An extremely interesting
application of the Doppler effect is to the discovery of
spectroscopic binaries. The Doppler principle has also been applied
to determine the nature of Saturn's ring. 

Since the effect of relative motion of source and observer
on apparent change in frequency or wavelength is common to all kinds
of waves. We expect that the de Broglie waves will also be Doppler shifted if there is a relative motion between the
emitter and the detector (see \cite{lee} for a very nice discussion). 
In this article, based on three old classic pieces of discoveries- the Doppler
effect in the year 1842, the Wien's displacement law in the year 1893 and the matter wave or the de Broglie wave in
the year 1923 \cite{weinberg,saha,landauQ}, we shall
study the Doppler shift of de Broglie waves associated with
fermions or bosons inside many body Fermi system or Bose system
respectively. In our investigation, instead of conventional photon or phonon gas, we
have considered bosons of non-zero mass. Then using the classic idea
of Wien we shall develop a formalism to express the variation of
de Broglie wavelength for the individual particles 
with the temperature of the system for both fermions and bosons. The
relations may be called as the modified form of the Wien's
displacement law for black body Fermi gas or Bose gas. To the best
of our knowledge this problem has not been addressed before.
\section{Doppler Shift of de Broglie Waves}
We consider a many body
quantum system consisting of either fermions or bosons. Then without
the loss of generality, we may assume that just like a black body system of electromagnetic waves or photon gas, the system is essentially a gas of
a large number of de Broglie waves of fermions or bosons. Hence we
may call the system as a
black body Fermi gas or Bose gas. We have further assumed that the
collisions among the particles is elastic in nature.
To obtain the change in
wavelength for de Broglie waves due to Doppler effect, we start with
the well known form of Lorentz transformations of particle momentum and
energy. We assume two frame of references, $S$, the rest frame and
the frame $S^\prime$ is moving with respect to $S$ with a uniform
velocity $V$ along $x$-direction. We further assume that the motion
of the particle is on $x-y$ plane. Then we have from the standard
text book results \cite{landauC}, the Lorentz transformation for the $x$-component
of particle momentum
\begin{equation}
p_x^\prime=\gamma\left ( p_x-\frac{VE}{c^2}\right ) ~~ {\rm{with}} ~~
\gamma=\left (1 -\frac{V^2}{c^2}\right )^{-1/2}
\end{equation}
where $E$ is the energy of the particle in $S$-frame and $c$ is the
velocity of light. If $\theta$ and $\theta^\prime$ are
the angles subtended by the particle momenta $\vec p$ and $\vec
{p^\prime}$ with the  $x$-direction in $S$
and $S^\prime$ frames respectively, then we have from eqn.(1)
\begin{equation}
p^\prime \cos\theta^\prime=\gamma \left( p\cos\theta -\frac{VE}{c^2}
\right ) 
\end{equation}
and since $p_y^\prime=p_y$,  hence we can write $p^\prime \sin\theta^\prime=
p\sin\theta$. 

Now defining the de Broglie wavelengths in these two frames as
$\lambda=h/p$ and $\lambda^\prime=h/p^\prime$, and using $E=(p^2c^2+E_0^2)^{1/2}$, the particle energy in $S$ frame
and a similar expression for $E^\prime$ in $S^\prime$ frame, with $E_0=m_0c^2$, the rest mass energy, we have
\begin{equation}
\frac{\lambda}{\lambda^\prime}\cos\theta^\prime =\gamma\left [
\cos\theta -\frac{V}{c} \left\{ 1+\left
(\frac{\lambda}{\lambda_c}\right )^2\right \}^{1/2}\right ]
\end{equation}
where $\lambda_c=h/m_0c$, the Compton wavelength. Then it can very 
easily be shown that the aberration is given by
\begin{equation}
\tan\theta^\prime=\frac{\sin\theta}{\gamma\left [
\cos\theta -\frac{V}{c} \left\{ 1+\left
(\frac{\lambda}{\lambda_c}\right )^2\right \}^{1/2}\right ] }
\end{equation} 
Now it is a matter of simple algebra to verify that for the mass-less case
when $\lambda_c=\infty $, the usual results for electromagnetic
waves or photons can be obtained from the eqns.(3) and (4).

To obtain the Doppler shift of the de Broglie waves for the particles it is 
more convenient to start from the Lorentz transformation for the particle energy,
given by
\begin{equation}
E^\prime=\gamma(E-Vp_x)
\end{equation}

\begin{widetext}
Squaring both the sides and using the definition of de Broglie waves in both
$S$ and $S^\prime$ frames, we have
\begin{equation}
\frac{\lambda}{\lambda^\prime}= \gamma\left[ 1-\frac{2V}{c}
\cos\theta \left\{ 1+\left (\frac{\lambda}{\lambda_c}\right )^2
\right\}^{1/2} +\left ( \frac{V}{c}\right )^2 \left \{ \cos^2\theta
+\left( \frac{\lambda}{\lambda_c}\right )^2 \right \} \right ]^{1/2}
\end{equation}
Again it is very easy to verify that the results for
electromagnetic waves follow from here for $\lambda_c=\infty$. 
\end{widetext}

It is also obvious
that the transverse form of Doppler shift with $\theta=\pi/2$ 
is non-vanishing in
the case of matter waves and is given by
\begin{equation}
\frac{\lambda}{\lambda^\prime}=\gamma\left [ 1+\left (
\frac{V}{c}\right )^2\left (\frac{\lambda}{\lambda_c}\right )^2\right
]^{1/2}
\end{equation}
\section{Variation of de Broglie Wavelength with Temperature} 
To obtain the variation of de Broglie wavelength with temperature,
we consider either a Fermi gas or a Bose
gas of non-zero mass in an enclosure, just like the
black body chamber of a photon gas. For the sake of simplicity the
enclosure is assumed to be spherical in nature. Further the wall of the
enclosure is assumed to be a perfect reflector and moving outward
adiabatically with a velocity $V$, where $V$ is small enough compared to the
velocity of light. The moving wall of the enclosure may be
treated as $S^\prime$ frame, whereas the $S$ frame is at rest inside
the enclosure with a fictitious observer sitting there. The tangential plane at
some arbitrary point on the outer surface of the wall is assumed to
be in the $y-z$ plane. Then the normal drawn from the centre to this
point of intersection is the $x$-direction. The particle which is hitting
the wall at this point of intersection is as before assumed to be moving in
$x-y$ plane. Then for a de Broglie wave of wavelength $\lambda$,
the point of intersection on the moving wall at which it is hitting is
equivalent to an observer moving away from the source, which is radially outward along $x$-direction. As a
consequence the received de Broglie wave at the moving wall will be red shifted and is
given by
\begin{equation}
\frac{\lambda}{\lambda^\prime} =\left [ 1-\frac{2V}{c}
\frac{\lambda}{\lambda_c}\cos \theta \right ]^{1/2}
\end{equation}
where we have neglected the term $(V/c)^2$ and assumed that $\lambda
\gg \lambda_c$ in the non-relativistic approximation. When the particle is reflected back from the point
of incidence, since the wall is moving outward, it is equivalent to
the emission from a source moving away from the observer. Therefore
in this case also the de Broglie wave of the particle will be
red shifted as observed from $S$ frame. Combining these two effects,
the relation between the final red shifted wavelength to that of
the original one is given by
\begin{equation}
\frac{\lambda^{\prime\prime}}{\lambda}=\frac{
\left[
1+\frac{2V}{c}\frac{\lambda^{\prime\prime}}{\lambda_c}\cos\theta\right
]^{1/2}}
{\left[ 1-\frac{2V}{c}\frac{\lambda}{\lambda_c}\cos\theta\right
]^{1/2}}
\end{equation}
Since $V/c \ll 1$, we have approximately
\begin{equation}
\frac{\lambda^{\prime\prime}}{\lambda}\approx \frac{
\left[ 1+\frac{V}{c}\frac{\lambda}{\lambda_c}\cos\theta\right ]}
{\left[ 1-\frac{V}{c}\frac{\lambda}{\lambda_c}\cos\theta\right ]}
\end{equation}
If we assume that the amount of final red shift is infinitesimal, the above
relation may further be approximated to 
\begin{equation}
\frac{\lambda^{\prime\prime}}{\lambda}\approx
1+\frac{2V}{c}\frac{\lambda}{\lambda_c}\cos\theta
\end{equation}
Writing the final red shifted wave length $\lambda^{\prime\prime}= \lambda
+d\lambda$, we have the resultant infinitesimal change in wave length
\begin{equation}
d\lambda=\frac{2V}{c} \frac{\lambda^2}{\lambda_c}\cos\theta
\end{equation}
To eliminate the arbitrary angle of incidence $\theta$, we consider multiple
reflection of de Broglie waves from the inner wall of the enclosure.
For spherical geometrical structure, with radius $r$, a de Broglie
wave travels a distance $2r\cos\theta$ between two successive
collisions. Therefore the number of reflections  per unit time is
$v/2r\cos\theta$. Hence the change in wavelength per unit time is 
\begin{equation}
d\lambda=\frac{V}{c}\frac{\lambda^2}{\lambda_c}\frac{dr}{r}
\end{equation}
where we have assumed that the particle travels $\delta r$ distance
in unit time and in the limiting case it is $dr$.

Now from the first law of thermodynamics we have for an adiabatic
change,
$dQ=dU+PdV_0=0$. Hence we know for a non-relativistic Fermi or Bose
gas $PV_0^{5/3}=$constant. 

\begin{widetext}
Now with the standard results from the textbook on
statistical mechanics \cite{huang}, the energy density for free Fermi gas or 
Bose gas is given by
\begin{equation}
\epsilon=\frac{3}{2}\frac{kT}{\lambda_0^3}f_{5/2}(z) ~~{\rm{and}} ~~
\epsilon=\frac{3}{2}\frac{kT}{\lambda_0^3}g_{5/2}(z) ~~
{\rm{respectively,}} ~~
\end{equation}
\[
{\rm{where}} ~~ 
f_{5/2}(z)=\sum_{l=1}^\infty (-1)^{l-1}\frac{z^l}{l^{5/2}}, 
~~ g_{5/2}(z)=\sum_{l=1}^\infty \frac{z^l}{l^{5/2}},
~~ \lambda_0=\frac{\hbar}{(2\pi m kT)^{1/2}} 
\]
\begin{equation} ~~{\rm{and}}~~ z=\exp\left
(\frac{\mu}{kT}\right )
\end{equation}
the fugacity, with $\mu$, the chemical potential of the constituents. The functions $f_\nu(z)$ and $g_\nu(z)$ are
the Fermi function and Bose function respectively.
\end{widetext}

Since for a many body Fermi system, whether it is electron gas in a 
piece of metal
or inside white dwarfs or neutron matter inside neutron stars, the
chemical potential $\mu$ is always non-zero. Therefore it is quite
obvious that from eqns.(14) and (15) an analytical
expression for energy density for a Fermi gas can not be obtained. We therefore approximate the Fermi gas by
Boltzmann statistics. Then it
is very easy to show that 
\[
P=\frac{2}{3}\epsilon ~~ {\rm{and}}~~ P \propto \exp\left (\frac{\mu}{kT}\right) T^{5/2}
\]
Hence from the adiabatic relation between pressure and volume, we have
\begin{equation}
\exp\left (\frac{2\mu}{5kT}\right) Tr^2=~~{\rm{constant}}
\end{equation}
where we have used $V_0=4\pi r^3/3$, the volume of the enclosure at
some instant. 
Taking log of both the sides and then differentiating all the terms
and finally using eqn.(13), we have
\begin{equation}
\left ( -\frac{2\mu}{5k}\frac{1}{T^2}+\frac{1}{T} \right )dT =-2
\frac{dr}{r}=-2\frac{c}{V}\frac{\lambda_c}{\lambda}
\frac{d\lambda}{\lambda}
\end{equation}

\begin{widetext}
Now integrating and rearranging the terms and finally redefining $\lambda$ and $T$ as
$\lambda^*$ and $T^*$, we have
\begin{equation}
\frac{1}{\lambda^*}-\frac{1}{T^*} =\ln\left (\frac{T}{T_0}\right )
\end{equation}
where 
\[
\frac{1}{\lambda^*}=\frac{2\lambda_c c}{V}\frac{1}{\lambda}, ~~
\frac{1}{T^*}=\frac{2\mu}{5k}\frac{1}{T} ~~ {\rm{and}}~~ T_0
~~{\rm{is~ a~ positive ~constant}}
\]
\end{widetext}

The above equation (eqn.(18)) looks like the equation for a thin lens, where the right
hand side is the inverse of focal length. Now it is well known that
at high temperature a Fermi gas behaves like a Boltzmann gas and then the
de Broglie wavelengths for the particles become infinitely large.
Therefore the right hand side must be negative, i.e., $T < T_0$.
Hence we can say that eqn.(18) may be compared with the equation for
the convex lens. The
temperature at which a fermion  behaves like a classical particle can be
obtained from the numerical solution of the 
equation $T^*\ln(T/T_0)+1=0$.
At this temperature, the de Broglie wavelength becomes infinitely
large and above this temperature eqn.(18) does not hold. The system
behaves classically. As
$T\longrightarrow 0$, the magnitude of the second term on the left hand side of
eqn.(18) becomes infinitely large much before the right hand side goes
to $-\infty$. Therefore we may conclude that as temperature
decreases, the de Broglie wavelength also decreases. Which means that the  quantum
mechanical effect becomes more and more important. In the extreme case, at $T\longrightarrow 0$, the Wave length 
tends to zero. If one compares eqn.(18) with the equation for a convex lens, then it is quite obvious that 
$\lambda=\infty$ corresponds to the object on the first focal plane of the convex lens in the object space, for which the 
real image is formed  at infinity. In the case of a Fermi gas
the temperature at which this happens, the de Broglie wavelength becomes infinitely large. This is also the
limiting temperature separating the temperature space into a quantum zone and a classical zone. 
Now it is well known that beyond the first focal plane away from the convex lens,
the images are always real, which in
the present scenario corresponds to classical picture. In the
case of a convex lens the object space between the focal plane and the
lens always produce virtual images. Same kind of picture is true here also. The
temperature zone between the upper critical value and
$T\longrightarrow 0$ is the quantum
mechanical region. So the quantum mechanical region of temperature in the present scenario
corresponds to the object space producing virtual images in the case of a convex lens. There is perhaps nothing
wrong in such comparison of quantum mechanical temperature zone with 
the object space produce virtual images. In the
quantum mechanical zone, because of uncertainty principle the exact location of the particle can not be predicted,
only the probability of existence at a point can be obtained from the wave function of the particle.
Therefore grossly speaking, a cloudy picture will be observed instead of a real location of the particle.

We next consider a massive Bose system. It may be a $\pi^+-\pi^-$ matter or
a neutral pion matter or even a system of extremely rarefied cold atoms. 
Since for these bosonic systems there is as such
no conserved quantum number, the chemical potential for the
constituents are exactly zero. Then we have from eqn.(18) (Now in the
case of bosons, the chemical potential $\mu=0$, therefore energy
density can be obtained from the second expression as given by
eqn.(14). The series $g_{5/2}(1)$ can be expressed in terms of known
Zeta-function)
\begin{equation}
\frac{1}{\lambda}=\frac{V}{2\lambda_c c} \ln\left (\frac{T}{T_0}
\right )~~ {\rm{or}}~~ \lambda\ln\left (\frac{T}{T_0}\right )
=~{\rm{constant}}
\end{equation}
This is again the form of Wien's displacement law for the massive Bos
gas.
From the nature of the above equations and the properties of massive
bosons, one can infer that $T_0$ is the minimum value of temperature 
for a Bose
system at which $\lambda=\infty$. In this crude model calculation we
may say that this is the temperature for Bose
condensation of the gas. Therefore, $T_0$ in eqn.(18) and eqn.(19) are 
carrying quite different physical meaning.
Since in the condensed phase all the bosons occupy the same quantum
state, the spatial coherence length will be large enough in the
atomic scale. In this simple model it is reflected by the extremely large 
value of de Broglie wavelength, which is large enough in the quantum
scale. Then as the temperature increases,
the system becomes more and more incoherent because of
the randomness, and at very high temperature the system becomes a
classical gas. Of course with this crude model calculation this can
not be shown.
For mass-less bosons, i.e. with $m=0$, since $\lambda_c=\infty$, the
above equation can not predict the condensation temperature. Which is already known for photon gas and phonon
gas.
\section{Conclusions}
It is therefore quite surprising that based
on three very old classic pieces of discoveries- the Doppler effect 
in the year 1842, the Wien's
displacement law in the year 1893 and the matter wave or the de Broglie wave
in the year 1923, it is possible to predict the variation of de
Broglie wavelength with temperature for 
individual fermions and bosons in a many
particle system. It is
also possible to obtain the temperature beyond which the fermionic
system behaves classically, the critical temperature for Bose
condensation for massive Bose gas and also one can conclude that for mass-less 
bosons the critical temperature of Bose condensation can not be predicted. 

\end{document}